\documentclass[english,a4paper,pre,twocolumn,amsmath,amssymb,longbibliography]{revtex4-1}
\usepackage{bm}
\usepackage{graphicx,color}
\usepackage{soul}
\usepackage{siunitx}
\usepackage{multirow}

\newcommand{\dd}{\mathrm{d}}

\newcommand{\mean}[1]{\langle #1 \rangle}

\newcommand{\IInt}[3]{\int_{#2}^{#3}\dd #1\;}

\renewcommand{\vec}[1]{\mathbf #1}

\newcommand{\sig}{\sigma}

\newcommand{\mev}{\milli\electronvolt}

\begin{document}

\title{Multiscale modelling of structure formation of C$_{60}$ on insulating CaF$_2$ substrates}

\author{William Janke}
\author{Thomas Speck}
\affiliation{Institut f\"ur Physik, Johannes Gutenberg-Universit\"at Mainz, Staudingerweg 7-9, 55128 Mainz, Germany}

\begin{abstract}
  Morphologies of adsorbed molecular films are of interest in a wide range of applications. To study the epitaxial growth of these systems in computer simulations requires access to long time and length scales and one typically resorts to kinetic Monte Carlo (KMC) simulations. However, KMC simulations require as input transition rates and their dependence on external parameters (such as temperature). Experimental data allows only limited and indirect access to these rates, and models are often oversimplified. Here we follow a bottom-up approach and aim to systematically construct all relevant rates for an example system that has shown interesting properties in experiments, buckminsterfullerene on a calcium fluoride substrate. We develop classical force fields (both atomistic and coarse-grained) and perform molecular dynamics simulations of the elementary transitions in order to derive explicit expressions for the transition rates with a minimal number of free parameters.
\end{abstract}

\maketitle


\section{Introduction}

The self-assembly of molecular building blocks has become a paradigm for the synthesis of novel materials with complex and hierarchical nanostructures~\cite{Whitesides2002}. In epitaxy experiments, surfaces are routinely exploited to facilitate and direct the growth of regular structures. Depending on the interactions of the deposited particles with the substrate different growth modes can be observed, leading to different morphologies of the growing films. In particular the growth processes of organic molecules on metallic~\cite{Barth07,Kuhnle09} and dielectric~\cite{HoffmannVogel2017} surfaces have sustained a high level of interest as a possible gateway to novel electronic devices manufactured at a molecular level. Tuning the subtle balance between molecule-molecule and molecule-surface interactions has opened up new pathways for creating an impressive variety of molecular structures on surfaces~\cite{Otero2011,Einax2013,Rahe2013,Kling2015}.

On the experimental side, much attention was given to the fullerene C$_{60}$ as a basis for carbon-based architectures. Studies have investigated the diffusion and film growth of C$_{60}$ on multiple metallic~\cite{Weckesser2001,Guo2004,Loske2009,Pawlak2011,Pawlak2012,Picone2016}, insulating~\cite{Chen1994,Chen2006,Szuba1999,Liu2006,Shin2010,Burke2007,LoskePaper,Groce2012,Rahe2012,Sato2017,Mitsuta2017,Seydel2018,Huttner2019,Nguyen2020,Guo2018}, and hybrid substrates~\cite{Tanigaki1993,Rossel2011,Matetskiy2013} resulting in a wide variety of cluster morphologies depending on substrate type, substrate temperature and particle flux. Despite these experimental successes, reliable structure prediction and generic design principles are still largely lacking. One route to further our understanding is computer simulations. Especially the kinetic Monte Carlo (KMC) method~\cite{KMCVoter2007} (or Gillespie algorithm~\cite{Gillespie76, Gillespie77}) has proven to be very promising since it is able to simulate the length and time scales needed to observe the cluster growth of these deposition experiments~\cite{LiuC60Sim,Korn11,Cantrell2012,bomm14,Kleppmann15,Kleppmann17,Acevedo16}.

The main challenge faced by KMC simulations is the modelling of all the possible elementary transition rates in the system, which commonly results in models with many free parameters or even thermodynamic inconsistencies. Experimental determination of the required input parameters (energy barriers and attempt rates) can be challenging~\cite{aeschlimann19} and such works are mostly focused on the determination of very few energy barriers. Therefore, the derivation of a complete KMC model directly from experimental data seems out of reach and the application of supporting computational methods is necessary. We have previously discussed the issues of common modelling approaches for KMC simulations~\cite{Janke2020} and introduced an approach for building a thermodynamically consistent model on the basis of molecular dynamics (MD) simulations of elementary transitions for the example of C$_{60}$ on C$_{60}(111)$ diffusion and film growth.

In this paper, we extend our approach to the epitaxial growth of C$_{60}$ on CaF$_2(111)$. This system undergoes molecular dewetting (transfer of molecules from the first layer of a cluster into higher layers before the completion of the monolayer~\cite{Burke2009}) and shows rich structural behaviour as temperature and coverage are varied~\cite{LoskePaper,Korn11}. There is a wide range of results on cluster densities, sizes, and morphologies available in the literature~\cite{LoskePaper,LoskeDiss}. In this paper, we are going to focus on the systematic derivation of a KMC model with a minimal set of free parameters, while the tuning and testing of the KMC simulations will be subject to a future publication. To achieve a comprehensive model, we derive force fields to be used in MD simulations, in which we observe different molecular configurations on the CaF$_2(111)$ surface for the measurement of several types of diffusive transition rates.

We start in Sec.~\ref{sec:modelling} with the introduction of the fullerene models and interaction potentials we are using for the MD simulations (implementation in LAMMPS~\cite{LAMMPSPlimpton}). Because of their spherical shape, it is often convenient to coarse-grain C$_{60}$ molecules into single beads with pairwise interaction potentials with other molecules and atoms. The effects of C$_{60}$ coarse-graining in MD simulations have been studied before~\cite{Abramo2004, Monticelli2012}, yielding relatively good agreement between coarse-grained and atomistic potentials in bulk for medium to high temperatures. However, for surface diffusion of C$_{60}$ on fine atomistic lattices like CaF$_2(111)$ it is easy to imagine that the molecular geometry and rotational degrees of freedom can have significant effects, especially as we are also interested in the lower temperature regime. We are therefore including a comparative study of a fully coarse-grained central body representation and an atomistic rigid-body representation for the C$_{60}$ molecule in this paper. The dynamics of the two models are compared in Sec.~\ref{sec:freediff} with a detailed analysis of the free diffusion process. Finally in Sec.~\ref{sec:edgediff} we look at transition types at the edge of a C$_{60}$ cluster and derive models for the observed transition rates with a single free parameter $\epsilon_\text{F}$.


\section{Building the model}
\label{sec:modelling}

\subsection{Substrate}

For the interaction potentials of the substrate atoms, we follow the example of Gillan's studies on CaF$_2$~\cite{Gillan1986}. The substrate is represented by point particles interacting pairwise through a Born-Mayer-Huggins style potential with an electrostatic contribution,
\begin{equation}
  \label{eq:sub}
  \phi^\text{S}_{ij}(r) = \frac{1}{4\pi\epsilon_0}\frac{q_iq_j}{r} + A_{ij}e^{-r/\rho_{ij}} - \frac{C_{ij}}{r^6}.
\end{equation}
The ions have charges $q_\text{Ca}=2e$ and $q_\text{F}=-e$ and the parameters of the repulsive and Van der Waals interaction are given in Tab.~\ref{tab:VdWParams}. A detailed discussion of these interaction parameters can be found in Ref.~\citenum{Gillan1986}.

\begin{figure}[t]
  \centering
  \includegraphics[width=3.2in]{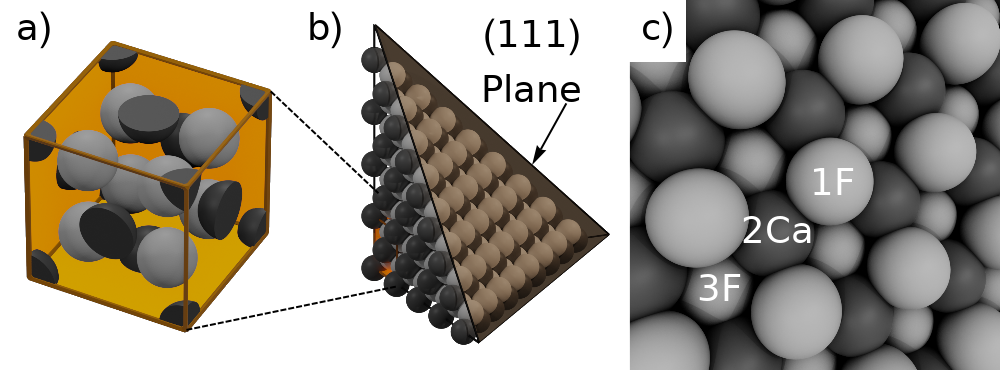}
  \caption{CaF$_2(111)$ structure. (a)~CaF$_2$ unit cell. (b)~Visualization of a CaF$_2$ crystal cleaved along the (111) plane. (c)~Close-up of the CaF$_2$(111) surface layer. Throughout, fluoride atoms are shown in light grey and calcium atoms in grey.}
  \label{fig:CaF2Structure}
\end{figure}

\begin{table}[b!]
  \centering
  \caption{Parameters taken from Ref.~\citenum{Gillan1986} for the repulsive and Van der Waals interactions of $\phi^\text{S}_{ij}$ [Eq.~\eqref{eq:sub}].}
  \label{tab:VdWParams}
  \begin{tabular}{c|c|c|c}
  \hline\hline
  & $A_{ij}[\SI{}{\electronvolt}]$ & $\rho_{ij}[\SI{}{\angstrom}]$ & $C_{ij}[\SI{}{\electronvolt\angstrom^6}]$ \\
  \hline
  F-F & $1808.0$ & $0.293$ & $109.1$ \\
  Ca-F & $674.3$ & $0.336$ & $0$ \\
  Ca-Ca & $0$ & --- & $0$ \\
  \hline\hline
  \end{tabular}
\end{table}

In all simulations, the substrate is composed of four layers of CaF$_2$ arranged in a fluorite structure [Fig. \ref{fig:CaF2Structure}(a)] and terminated by a layer of fluoride in the (111) plane [Fig. \ref{fig:CaF2Structure}(b)]. A close-up of the surface of CaF$_2$(111) is shown in Fig.~\ref{fig:CaF2Structure}(c). Each layer is itself composed of three sublayers of atoms organized in hexagonal lattices on different heights. The top (1F) and bottom (3F) sublayers are filled with fluoride atoms with a sublayer of calcium (2Ca) in between. Atoms are initialized on the corresponding lattice positions and equilibrated at a desired temperature using a Langevin thermostat to achieve random initial velocities, followed by velocity rescaling. Atoms in the bottommost CaF$_2$ layer are kept fixed throughout and do not move.

\subsection{Atomistic fullerene model}

In the atomistic fullerene model, we neglect molecular vibrations and represent each C$_{60}$ molecule by 60 carbon atoms that are grouped together as a rigid body. Single carbon atoms of different molecules interact through the standard Lennard-Jones potential
\begin{equation}
  \phi_\text{C}(r) = 4\epsilon_\text{C}\left[\left(\frac{\sig_\text{C}}{r}\right)^{12}-\left(\frac{\sig_\text{C}}{r}\right)^6\right].
\end{equation}
To determine the parameters $\epsilon_\text{C}$ and $\sig_\text{C}$, we perform MD simulations of $19$ C$_{60}$ molecules arranged in a hexagonal cluster on the CaF$_2$($111$) substrate. The parameters were then tuned to reach an average potential energy per lateral neighbour ("bond energy") of $\SI{270}{\mev}$ and an average center-to-center distance between nearest neighbours of $\SI{10.1}{\angstrom}$ at $T=\SI{200}{\kelvin}$, matching the cluster properties of the Girifalco potential (Sec. \ref{subsec:CGModel}) on CaF$_2$($111)$. The results of this parameter tuning are $\epsilon_\text{C}=\SI{2.36}{\mev}$ and $\sig_\text{C}=\SI{3.62}{\angstrom}$.

For the interactions between the individual carbon and substrate atoms, we assume a standard Lennard-Jones C-F 
\begin{equation}
  \phi_\text{F}(r) = 4\epsilon_\text{F}\left[\left(\frac{\sig_\text{F}}{r}\right)^{12}-\left(\frac{\sig_\text{F}}{r}\right)^6\right]
  \label{eq:atomisticF}
\end{equation}
and a Buckingham C-Ca interaction 
\begin{equation}
  \phi_\text{Ca}(r) = Ae^{-r/\rho}
  \label{eq:atomisticCa}    
\end{equation}
inspired by the interaction potentials obtained in Ref.~\citenum{CopperLeeuwCaF2CPots2003}. In this reference, various interatomic interaction potentials are listed for CaF$_2$ with several different molecules. While C$_{60}$ was not part of this study, the listed parameters of CaF$_2$ with oxygen and carbon atoms of different molecules are taken as reference points. The length scale parameters are chosen as $\rho=\SI{0.297}{\angstrom}$ based on the given Ca-O and Ca-Ow potentials, and as $\sig_\text{F}=\SI{2.055}{\angstrom}$, based on the CD-F interaction. The repulsive parameter is chosen as $A=\SI{1300}{\electronvolt}$ to be in the same range as the listed Ca-Ow, Ca-F and Ca-O potentials. Arguably the most important parameter is $\epsilon_\text{F}$, as it determines the attractive interaction of carbon and fluoride atoms and consequently governs the overall interaction strength of the C$_{60}$ molecule with the CaF$_2$ substrate. We are therefore not estimating this parameter and instead leave it as a free variable that we are going to vary in the range $\epsilon_\text{F}\in[35,55]\;\SI{}{\mev}$. This range results in a total molecule-substrate binding energy of $300-\SI{800}{\mev}$. We expect the dewetting barrier for C$_{60}$ on CaF$_2$($111$) to be in this range since C$_{60}$ has been observed to grow into two-layered clusters on CaF$_2$($111$)~\cite{LoskePaper,Korn11}, suggesting a molecule-substrate interaction that is somewhat weaker than the molecule-molecule interaction (the total bond energy of C$_{60}$ on C$_{60}$(111) is about $\SI{900}{\mev}$).

\subsection{Coarse-grained fullerene model}
\label{subsec:CGModel}

Going a step further, the rigid model of C$_{60}$ is coarse-grained into a single point particle (a spherical ``bead''). These beads interact pairwise through the well-known Girifalco potential~\cite{GirifalcoPot,GirifalcoPot92}
\begin{multline}
  \nonumber\phi^\text{CG}_{\text{C}_{60}}(s) = -\alpha\left(\frac{1}{s(s-1)^3}+\frac{1}{s(s+1)^3}-\frac{2}{s^4}\right) \\
  + \beta\left(\frac{1}{s(s-1)^9}+\frac{1}{s(s+1)^9}-\frac{2}{s^{10}}\right)
\end{multline}
with two parameters, $\alpha=\SI{46.7e-3}{\electronvolt}$ and $\beta=\SI{84.5e-6}{\electronvolt}$. The variable $s=r/R$ is the center-to-center distance of two interacting molecules $r$, scaled by the nucleus-to-nucleus diameter of C$_{60}$, $R=\SI{7.1}{\angstrom}$. Because the coarse-grained representation has lost the rotational degrees of freedom of the C$_{60}$ molecules, it is usually assumed to be a good approximation for temperatures well above $\SI{260}{\kelvin}$. At this temperature, crystalline C$_{60}$ undergoes a structural phase transition from a plastic crystal, in which orientations are disordered (at high temperatures), to a phase in which the molecular orientations align (at low temperatures)~\cite{David1992,Moret1993,Kasatani1993,Yoneda1997,Bozhko2011,Bozhko2015}. While the coarse-grained representation was therefore mostly used for studies at high temperatures -- like the predictions of a stable liquid phase of C$_{60}$ \cite{Hagen1993,Caccamo1997,Hasegawa1999,Fartaria2002,Cheng1993} or the examination of C$_{60}$/C$_{70}$ mixtures \cite{Knia1995,Khusnutdinoff2015} -- the simplicity of the potential occasionally motivates its use at lower temperatures~\cite{Rey1994,Jin2015,Royall2011,Hu2019,LiuC60Sim,Janke2020}. A comparative study between coarse-grained and atomistic versions of this potential was previously done for C$_{60}$ in bulk~\cite{Abramo2004}.

\begin{figure}[t]
  \includegraphics[width=3.2in]{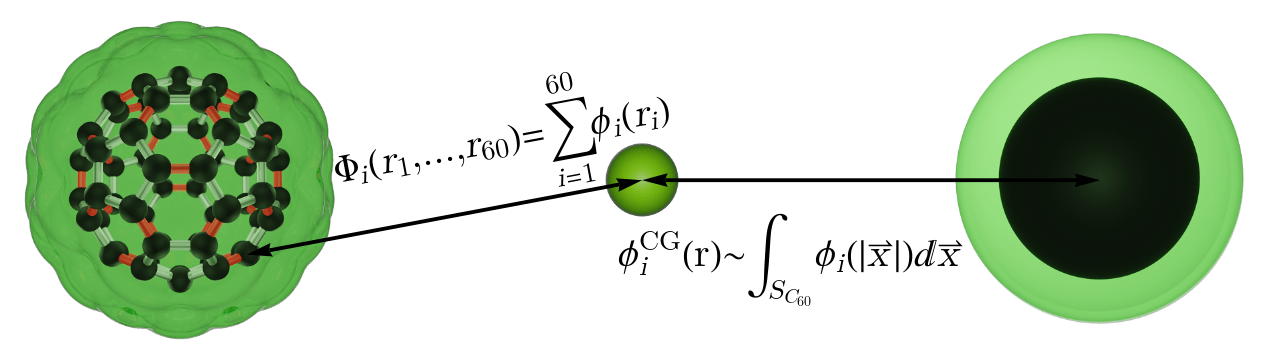}
  \caption{Atomistic (left) \emph{vs.} coarse-grained (right) interaction of a C$_{60}$ molecule with a single atom.}
  \label{fig:cg}
\end{figure}

To derive coarse-grained potentials for the interaction of C$_{60}$ molecules with the CaF$_2$ substrate atoms, we follow an approach similar to the derivation of the Girifalco potential (and as it was previously also done in other works \cite{Breton1993,Abramo1997,Palucha2002}) by smearing out the C atoms over a sphere of radius $R=\SI{7.1}{\angstrom}$ and integrating the atomistic interactions (\ref{eq:atomisticF}) and (\ref{eq:atomisticCa}) over this sphere, cf. Fig~\ref{fig:cg}. Written in spherical coordinates, the coarse-grained potentials are obtained from
\begin{equation}
  \phi^\text{CG}_i(r) = -2\pi R^2\eta\IInt{\theta}{0}{\pi}\phi_i(\sqrt{R^2+r^2-2rR\cos\theta}),
\end{equation}
where the subscript $i$ is either F or Ca, and $\eta=60/(4\pi R^2)$ is the number density of carbon atoms on the sphere. As a result, we obtain the coarse-grained potentials
\begin{multline}
  \phi_\text{F}^\text{CG}(r) = \frac{60}{Rr}\epsilon_\text{F}\biggl(\frac{\sigma_\text{F}^6}{2(R+r)^4}-\frac{\sigma_\text{F}^{12}}{5(R+r)^{10}}\\
  -\frac{\sigma_\text{F}^6}{2(R-r)^4}+\frac{\sigma_\text{F}^{12}}{5(R-r)^{10}}\biggr)
\end{multline}
and
\begin{multline}
  \phi_\text{Ca}^\text{CG}(r) = \frac{60A\rho^2}{Rr}e^{-r/\rho}\biggl[\sinh{\left(\frac{R}{\rho}\right)}\left(1-\frac{R}{\rho}\right)\\
  +\frac{r}{\rho}\cosh{\left(\frac{R}{\rho}\right)}\biggr].    
\end{multline}
The parameters $\epsilon_\text{F}$, $\sigma_\text{F}$, $A$, and $\rho$ are the same as in the atomistic model (with $\epsilon_\text{F}$ as a free parameter).


\section{Free diffusion}
\label{sec:freediff}

\begin{figure}[b]
  \includegraphics[width=3.2in]{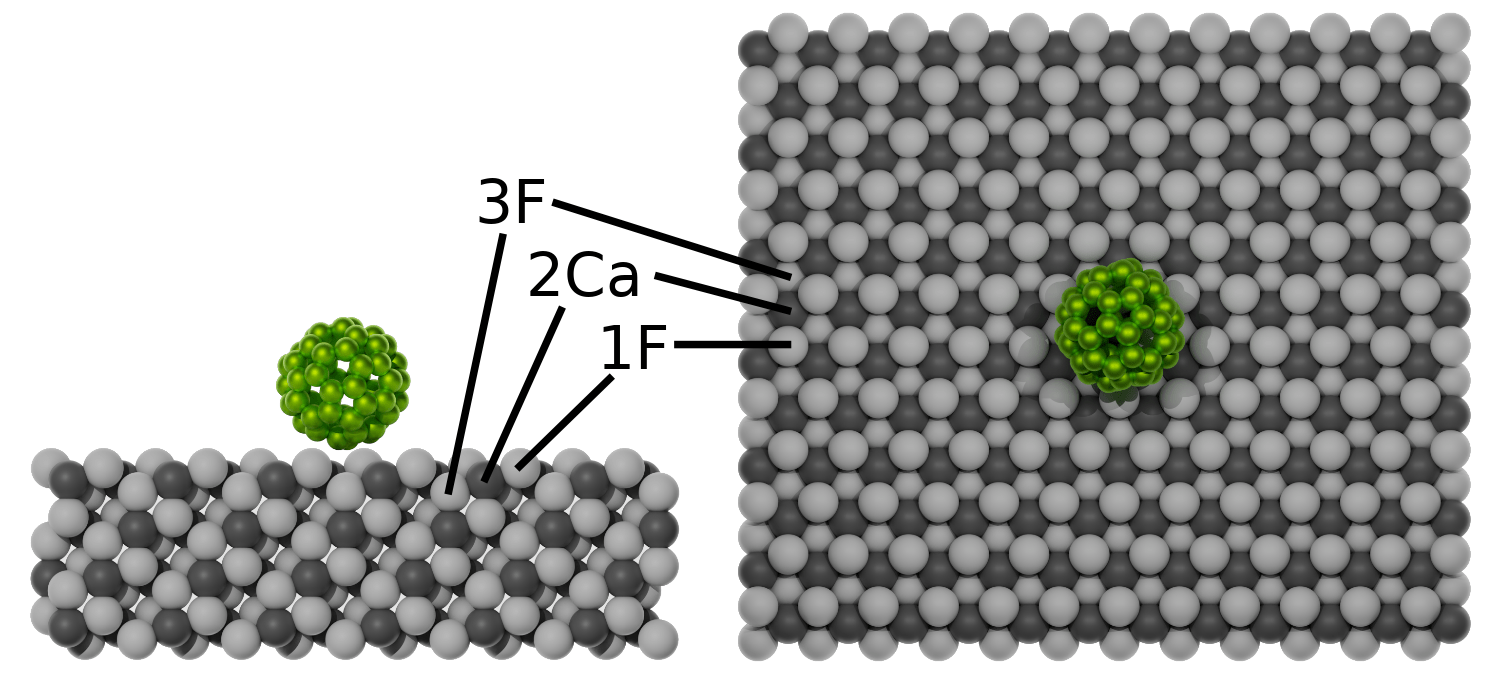}
  \caption{Configuration for MD simulations of the free diffusion process of C$_{60}$ on a CaF$_2(111)$ surface. Fluoride atoms are represented as light grey, calcium atoms as grey and carbon atoms as green spheres.}
  \label{fig:FreeDiffConf}
\end{figure}

We first study the diffusive motion of a single molecule, whereby we are particularly interested in comparing the atomistic rigid model with the coarse-grained model. The MD simulations in this section are set up in a box approximately sized $\SI{40}{\angstrom}\times\SI{46}{\angstrom}$ with periodic boundary conditions in the $x$ and $y$ direction. A single C$_{60}$ molecule is placed on top of the substrate as shown in Fig.~\ref{fig:FreeDiffConf}.

At the beginning of the simulation, a combination of Langevin thermostat and velocity rescaling is used to achieve random initial velocities at the desired temperature. After a short period of equilibration, a $\SI{50}{\nano\second}$ trajectory of the C$_{60}$ molecule is recorded for later analysis. For each temperature ($T\in [40,450]\;\SI{}{\kelvin}$, covering low to medium high temperatures) and interaction strength ($\epsilon_\text{F}\in\{35,40,45,50,55\}\SI{}{\milli\electronvolt}$), ten of these simulations are run, totalling $\SI{500}{\nano\second}$ of diffusion per set of parameters.

\subsection{Transition pathways}
\label{sec:MEPs}

In the MD trajectories, we see for both the atomistic and coarse-grained model that the C$_{60}$ molecules preferably occupy positions above third-layer fluoride atoms (position 3F in Fig. \ref{fig:FreeDiffConf}), which we identify as lattice sites. Before discussing the MD simulation results, we take a look at the minimum energy paths (MEPs) to gain insight into the typical transition path a C$_{60}$ molecule takes from one lattice site to another. We determine the MEP of a C$_{60}$ molecule on a stationary CaF$_2$(111) surface using the drag- and NEB-method~\cite{Henkelman2002}, which we implemented in Mathematica~\cite{Mathematica}. For the coarse-grained model, we take a straight line connecting two neighbouring lattice sites as an initial path for the NEB method to optimize. Because of the rotational degrees of freedom, in the atomistic model it is less trivial to construct an initial path. Therefore, we first calculate a MEP using the drag method and then further optimize it with the NEB method. The resulting MEPs and the corresponding potential energies along the MEP are shown in Fig.~\ref{fig:MEPs} (animations of the transition paths are provided in the Supplementary Information~\cite{sm}). The energy minima ($E_\text{min}$), maxima ($E_\text{max}$), and barriers ($\Delta E_\text{D}$) from these paths are summarized in Tab.~\ref{tab:MEPResults}.

\begin{figure}
  \includegraphics[width=3.2in]{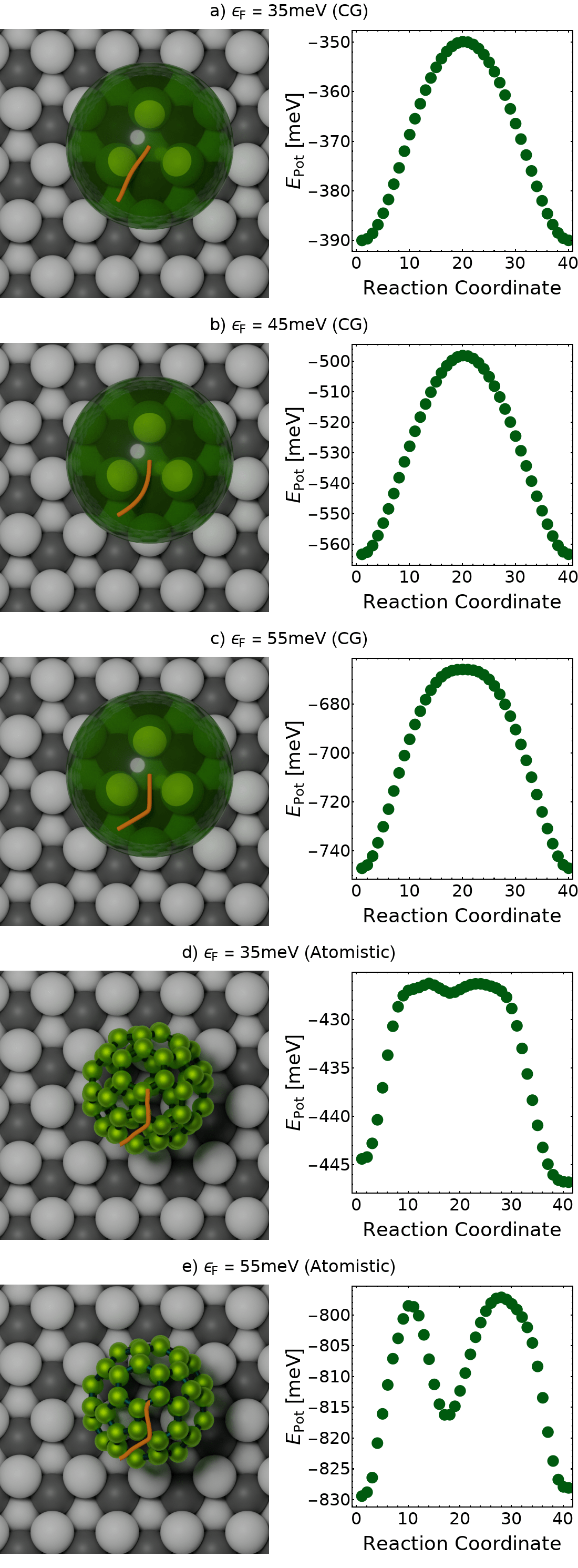}
  \caption{Minimum energy paths (MEPs) for surface diffusion in (a-c)~the coarse-grained and (d-e)~the atomistic model for several values of $\epsilon_\text{F}$. Left: Initial configuration with the center-of-mass trajectory of the MEP in orange. Right: Potential energy along the MEP.}
  \label{fig:MEPs}
\end{figure}

\begin{table}[b]
  \centering
  \caption{Minimum energy path (MEP) analysis results. All values are given in $\SI{}{\milli\electronvolt}$}
  \label{tab:MEPResults}
  \begin{tabular}{c|ccc|ccc}
  \hline\hline
  & \multicolumn{3}{c}{Atomistic} & \multicolumn{3}{c}{Coarse-Grained} \\
  \hline
  $\epsilon_\text{F}$ & $E_\text{min}$ & $E_\text{max}$ & $\Delta E_\text{D}$ & $E_\text{min}$ & $E_\text{max}$ & $\Delta E_\text{D}$ \\
  \hline
  $35$ & $-447$ & $-426$ & $20.5$ & $-390$ & $-350$ & $40.1$ \\
  $40$ & $-537$ & $-516$ & $21.2$ & $-475$ & $-421$ & $54.0$ \\
  $45$ & $-631$ & $-608$ & $22.8$ & $-563$ & $-498$ & $65.1$ \\
  $50$ & $-728$ & $-702$ & $25.8$ & $-654$ & $-580$ & $73.8$ \\
  $55$ & $-829$ & $-797$ & $32.3$ & $-747$ & $-666$ & $81.2$ \\
  \hline\hline
  \end{tabular}
\end{table}

\begin{figure*}[t]
  \includegraphics[width=6.4in]{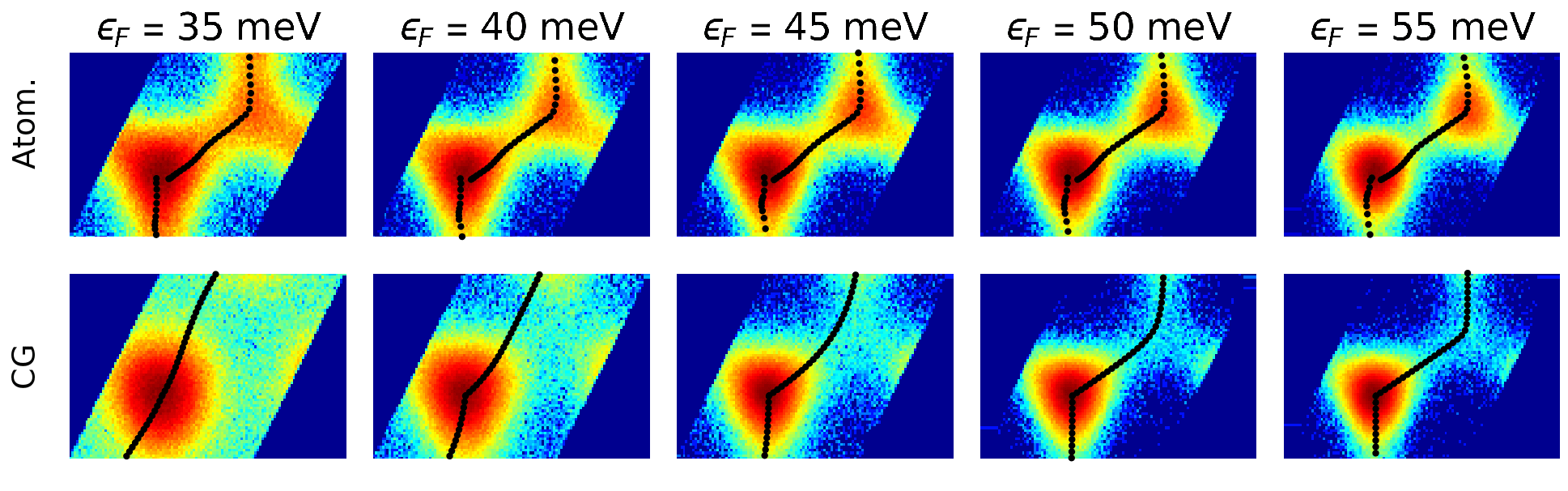}
  \caption{Density plots of $\ln p(x,y)$ at $T=\SI{100}{\kelvin}$ with the minimum energy paths (MEPs) overlaid as black dots for the atomistic and coarse-grained (CG) model at several values of $\epsilon_\text{F}$.}
  \label{fig:allDensPlots}
\end{figure*}

The MEPs of the two models differ in several ways. Firstly, the value of the bond energy between the atomistic molecule and the substrate is $10-20\%$ larger. This is to be expected due to the atomistic model's ability to align the molecular orientation with respect to the substrate atom positions. Secondly, because the atomistic molecule can also change its alignment during diffusive transitions, the diffusion barrier is significantly lower than in the coarse-grained model. This rotational mobility causes the atomistic model to have an additional local energy minimum between the two lattice sites on top of the surface calcium. This energy minimum is very weak for $\epsilon_\text{F}=\SI{35}{\milli\electronvolt}$ but gets stronger with increasing interaction strength since the C-F interaction overwhelms the repulsion of the Ca atoms. It also causes the atomistic model to always pass over the surface calcium (Fig.~\ref{fig:MEPs}). In contrast, the coarse-grained model tends to take a more direct path between the lattice sites at $\epsilon_\text{F}=\SI{35}{\milli\electronvolt}$ and only passes over the calcium at stronger interaction strengths.

\subsection{Energy landscapes}
\label{sec:PPD}

The potential energy plots of the previous section give an idea of how the potential energy landscapes of our models look like. In this section, we extract the free energy landscapes from our MD data at non-zero temperatures to see if they match with the expectations established by the MEPs. To this end, we determine the positional probability distribution $p(x,y)$ by projecting the center-of-mass trajectories into a single unit cell.

\begin{figure}[b!]
  \centering
  \includegraphics[width=3.2in]{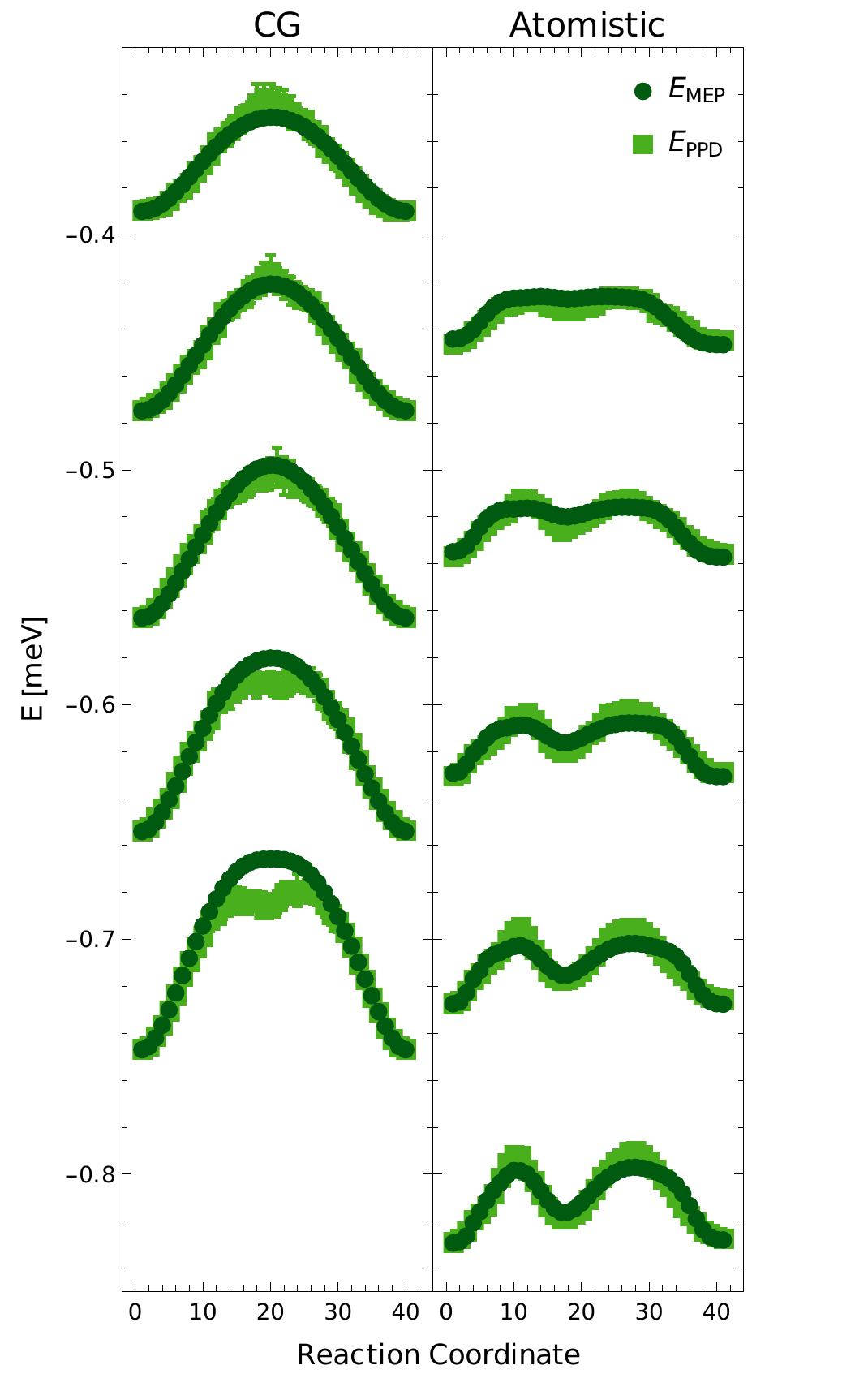}
  \caption{Comparison of the potential energies $E_\text{MEP}$ along the minimum energy paths (MEPs) from the previous section to the free energies $E_\text{PPD}$ (extrapolated from MD data to $\SI{0}{\kelvin}$) along the MEPs. The binding energy $\epsilon_\text{F}$ is increased from $\epsilon_\text{F}=\SI{35}{\milli\electronvolt}$ (top) to $\epsilon_\text{F}=\SI{55}{\milli\electronvolt}$ (bottom) in steps of $\SI{5}{\milli\electronvolt}$.}
  \label{fig:MEPvsPPD}
\end{figure}

From the positional probability distribution, we can define a free energy landscape through
\begin{equation}
  F(x,y;T) = -k_\text{B}T\ln p(x,y;T)
  \label{eq:FreeEn}
\end{equation}
up to a constant offset. For an easier comparison of the MEP results with the free energy landscapes, we set this constant offset to match the potential energy minimum of the MEPs. Qualitatively, we can already note here that the MEPs align nicely with the obtained free energy landscapes (Fig.~\ref{fig:allDensPlots}). We can go a step further and extract an effective substrate potential $E(x,y)$ by exploiting that the free energy can be written as $F(x,y;T)=E(x,y)-TS(x,y)$ with positional entropy $S(x,y)$, which we assume to be independent of temperature. In Fig.~\ref{fig:MEPvsPPD}, the resulting potential energies are plotted in comparison to the potential energy curves along the MEPs of the previous section. We obtain a very good agreement between the energy curves obtained by these two different approaches.

\subsection{Diffusion coefficients}
\label{sec:MSD}

A common approach for determining energy barriers and attempt rates of diffusive processes is to determine the translational diffusion coefficient $D(T)$ from the time-dependence of the mean squared displacement (MSD),
\begin{align}
  \text{MSD}(t)=\mean{[\vec{x}(t)-\vec{x}(0)]^2}=2fDt,
\end{align} 
with $f=2$ being the dimensionality of the diffusion process on a surface. Assuming that the diffusion process can be described by a random walk with Arrhenius type waiting times between diffusive transitions, we can put $D$ in relation with an attempt rate $\nu_0$ and an energy barrier $\Delta E_\text{D}$,
\begin{align}
  D=\frac{1}{2f}\nu_0\mean{l^2}\exp{\left(-\frac{\Delta E_\text{D}}{k_BT}\right)},
  \label{eq:ArrheniusDiff}
\end{align}
where $\mean{l^2}$ is the mean squared jump length of the diffusion process, which would be equal to $\mean{l^2}=0.386^2\SI{}{\nano\metre^2}=\SI{0.149}{\nano\metre^2}$ if we assume only single jumps between neighbouring lattice sites. However, this assumption is not accurate in our case since in the simulations we observe jumps across multiple lattice sites, especially for higher temperatures. The coarse-grained model occasionally even shows long diffusive jumps without a change in direction, reminiscent of L\'evy flights \cite{Klafter1994,Klafter1996}.

\begin{figure}[t]
\includegraphics[width=3.2in]{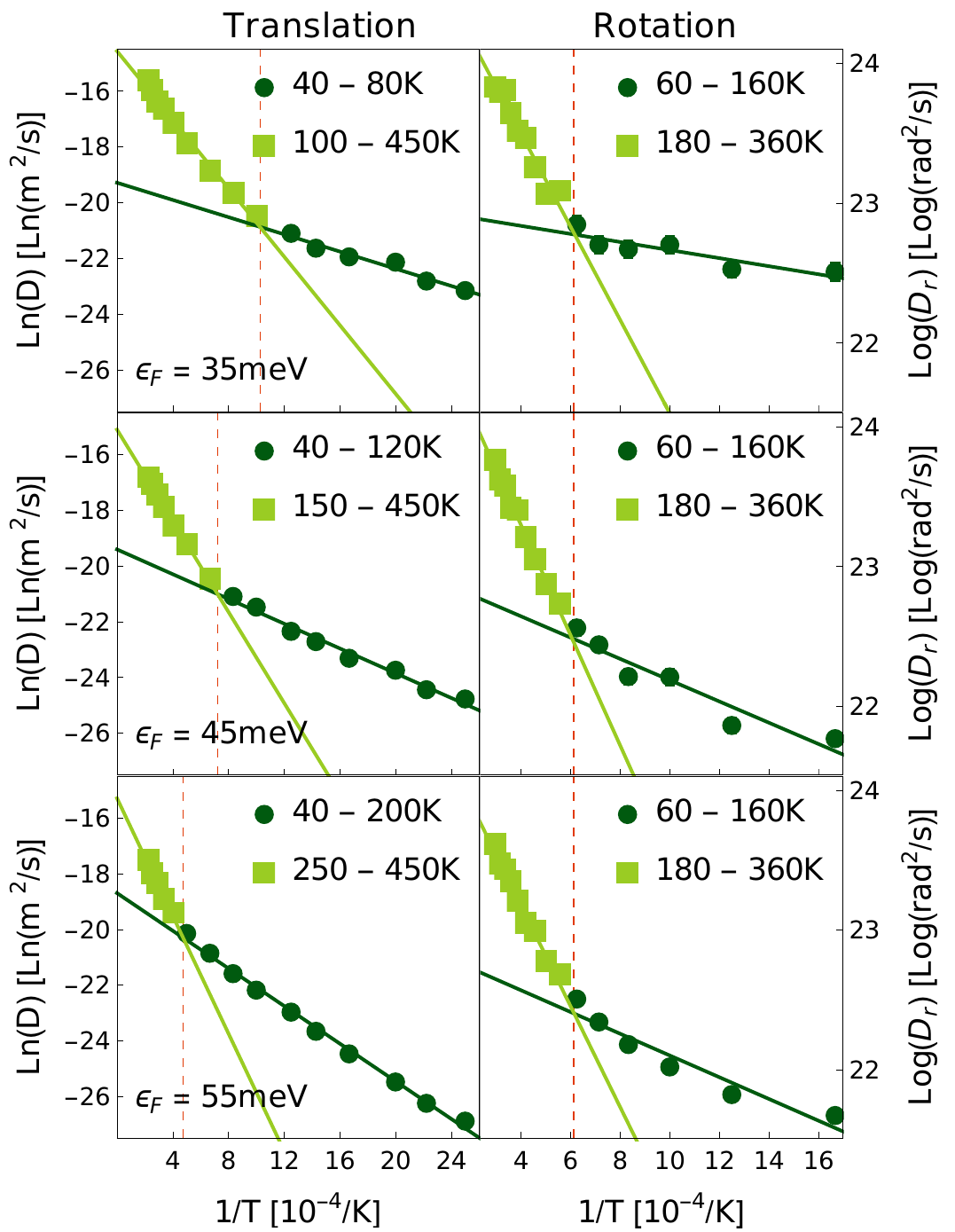}
\caption{Crossover in translational and rotational diffusive behaviour seen in the atomistic model. The dashed red lines mark the crossover temperatures.}
\label{fig:rotDiffCrossover}
\end{figure}

While the coarse-grained model always exhibits an Arrhenius-type behaviour (that is, the data points in a $\ln D$ \emph{vs.} $1/T$ plot fall on a line, see Supplementary Information for a comparison of the Arrhenius data of both models~\cite{sm}), the atomistic model shows a crossover in diffusive behaviour somewhere in the range of $100-\SI{200}{\kelvin}$, depending on the value of $\epsilon_\text{F}$ (Fig.~\ref{fig:rotDiffCrossover} left). Because this temperature range hints at the crossover being related to the structural phase transitions of C$_{60}$ \cite{David1992,Moret1993,Kasatani1993,Yoneda1997,Bozhko2011,Bozhko2015}, we look at a few additional shorter simulations with a higher time resolution of the output trajectory to analyse the rotational diffusion of the atomistic model. These simulations are run for $\epsilon_\text{F}\in\{35,45,55\}\SI{}{\milli\electronvolt}$ at temperatures $T\in[60,\SI{360}{\kelvin}]$ for a total of $\SI{50}{ns}$ per set of parameters. For details on how to extract the rotational diffusion coefficient, see appendix~\ref{app:rotDiff}. The results are shown in the Arrhenius plots in Fig.~\ref{fig:rotDiffCrossover}. The crossover is also clearly visible in the rotational diffusive behaviour and appears consistently at a temperature of around $\SI{163(1)}{\kelvin}$. 

The temperature range of the crossover we observe in the translational and rotational diffusion coefficients are lower than the temperatures at which the rotational phase transition of C$_{60}$ in thin films was observed experimentally \cite{Bozhko2011,Yoneda1997} ($220-\SI{260}{\kelvin}$). Deposition experiments of C$_{60}$ on a metal-silicon surface have also found a crossover in diffusive behaviour in a temperature range of $140-\SI{160}{\kelvin}$~\cite{Matetskiy2013}. In MD simulations of the diffusion process of C$_{60}$ on graphene \cite{JafaryZadeh2012}, a very similar observation of a diffusive crossover was made in a temperature range of $25-\SI{75}{\kelvin}$. In that system, the energy landscape on graphene yields much smaller energy barriers than in our case, shifting the crossover to a very low temperature. We can conclude that our observation of a crossover temperature that changes with the interaction strength $\epsilon_\text{F}$ is in agreement with the variation of experimentally observed crossover temperatures on different substrates.

\subsection{Modelling the diffusion barrier}

\begin{figure}[t]
  \includegraphics[width=3.2in]{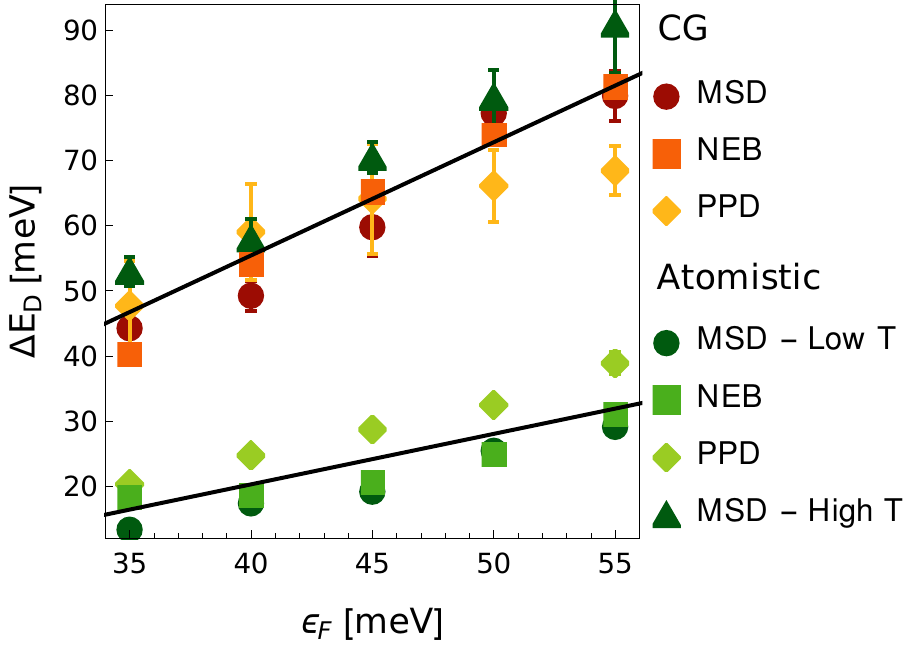}
  \caption{Summary of energy barriers for the free diffusion process. Linear fits to the two sets are shown in black. MSD: Energy barriers obtained from the mean-squared displacement analysis in Sec.~\ref{sec:MSD}. NEB: Energy barriers taken from the minimum energy paths of the nudged elastic band method in Sec.~\ref{sec:MEPs}. PPD: Energy barriers obtained from the energy landscapes calculated from the positional probability distributions in Sec.~\ref{sec:PPD}.}
  \label{fig:enBarrSummary}
\end{figure}

Exploiting Eq.~\eqref{eq:ArrheniusDiff}, we determine energy barriers and attempt rates from the MSD analysis for the coarse-grained and atomistic model. The resulting energy barriers are summarized in Fig. \ref{fig:enBarrSummary} in conjunction with the energy barriers we have previously determined from the transition pathways (NEB) and energy landscape analysis (PPD). The obtained values can be grouped into two sets. The first set of lower energy barriers ($14-\SI{40}{\milli\electronvolt}$) comes from the low temperature analyses of the atomistic model and is therefore likely an accurate set of energy barriers for the diffusion process at low temperatures. The second set contains higher energy barriers ($40-\SI{90}{\milli\electronvolt}$) and consists of the analyses on the coarse-grained model as well as the high temperature MSD results of the atomistic model. Linear fits on these two sets of energy barriers yield the two models,
\begin{align}
  \Delta E_\text{D,LT}&=\SI{-10.7(67)}{\milli\electronvolt}+0.78(14)\cdot\epsilon_\text{F} \label{eq:EdLT} \\
  \Delta E_\text{D,HT}&=\SI{-14.1(82)}{\milli\electronvolt}+1.74(18)\cdot\epsilon_\text{F},\label{eq:EdHT}
\end{align}
for the low temperature (LT) and high temperature (HT) regimes, respectively.

\subsection{Discussion}

An experimental value for the diffusion barrier of C$_{60}$ on CaF$_2(111)$ was determined by Felix Loske \emph{et al.} through the observation of cluster densities in a molecular beam epitaxy experiment in a temperature range $96-\SI{217}{\kelvin}$~\cite{LoskePaper}. They obtained a value of $\Delta E_\text{D}=214(16)\SI{}{\milli\electronvolt}$, which is in strong disagreement with our results in both temperature regimes. Extrapolating from Eqs. (\ref{eq:EdLT}) and (\ref{eq:EdHT}), we can achieve such an energy barrier for values of
\begin{align*}
  \epsilon_\text{F,Exp,LT}=\SI{288(56)}{\milli\electronvolt} \hspace{10px}
  \epsilon_\text{F,Exp,HT}=\SI{131(17)}{\milli\electronvolt}.
\end{align*}
However, at such high values of $\epsilon_\text{F}$, the adparticle-substrate interaction would be much stronger than the adparticle-adparticle interaction (with dewetting barriers greater than $\SI{2}{\electronvolt}$), which seems at odds with the observation of the two-layered cluster growth of C$_{60}$ on CaF$_2$ \cite{Korn11,LoskeDiss}. A probable explanation for this discrepancy was mentioned in Ref.~\citenum{Janke2020Erratum}: Small impurities on the substrate surface could strongly affect the diffusion process if they act as nucleation sites, leading to an effective diffusion process with long jumps between impurities. The result of such a process can naturally not be compared to the diffusion of C$_{60}$ on a clean, defect-free CaF$_2$ substrate as we have simulated here.

Another possible explanation could be the effect of polarizability of the C$_{60}$ molecule. It was previously shown that it can have a significant impact on the growth of C$_{60}$ films on ZnPc/AG$(111)$, where chain phases were observed \cite{Jin2015}. Considering that the top fluoride layer of CaF$_2(111)$ is negatively charged, it can be suspected that polarizability may play a role in this diffusion process as well. Because of the discrepancy in the energy barriers we are not going to discuss the attempt rates in detail and just note that they are of the order $\SI{e11}{\hertz}$ in the low temperature regime of the atomistic model, $\SI{e12}{}-\SI{e13}{\hertz}$ in the high temperature regime of the atomistic model, and $\SI{e13}{}-\SI{e14}{\hertz}$ for the coarse-grained model (see Supplementary Information~\cite{sm}). These attempt rates are also in disagreement with the attempt rate from the cluster growth experiment, $\nu_0=\SI{3.2e17}{\hertz}$ \cite{Janke2020Erratum}. However, a range of other experiments with large molecules report attempt rates in the range of $\SI{e10}{}-\SI{e14}{\hertz}$~\cite{Weckesser1999,Weckesser2001,Schunack2002} compatible with our simulations.

To conclude this section, analysing in detail the diffusion process of the atomistic and coarse-grained model using three different approaches, we have found that they exhibit different behaviour at low temperatures. The difference seems to diminish at higher temperatures according to our mean-squared displacement analysis. This observation suggests that the coarse-grained model should only be used at higher temperatures, whereas the atomistic model probably yields better results for the low temperature regime. To be able to test this conclusion, we are going to measure edge diffusion transition rates for both the coarse-grained and atomistic model in the following section. The accurate modelling of edge diffusion transitions then enables a future study involving KMC simulations to compare the resulting cluster morphologies of the two models with experimental observations.


\section{Edge diffusion}
\label{sec:edgediff}

\begin{figure*}[t!]
  \includegraphics[width=\textwidth]{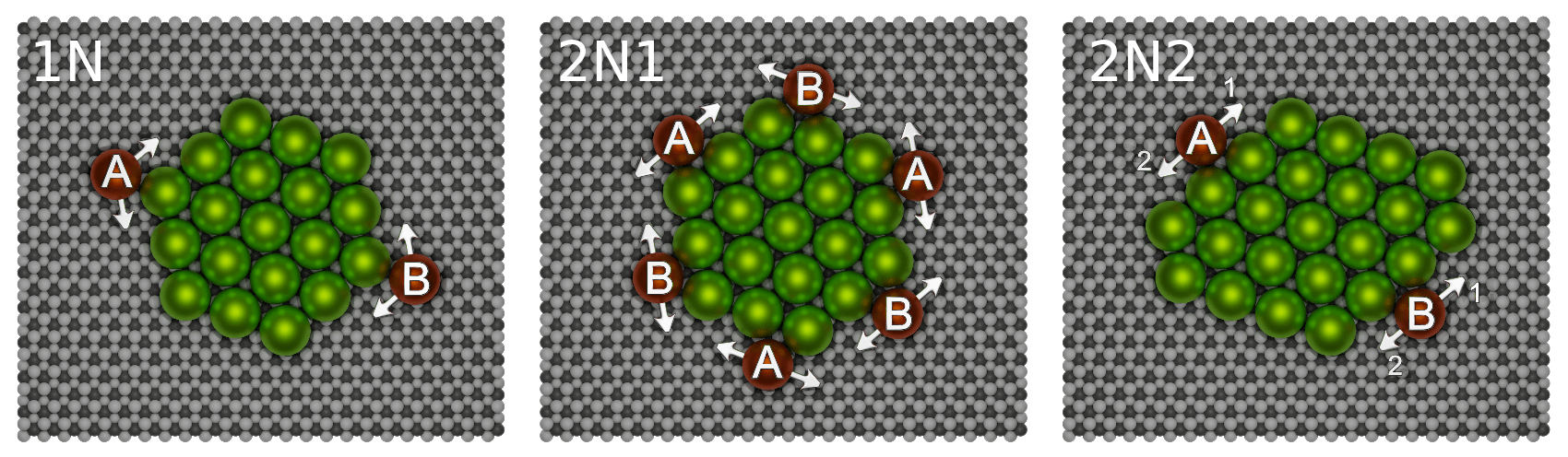}
  \caption{Initial configurations for edge diffusion with one and two initial neighbours. Particles that can undergo transitions of interest are highlighted in orange.}
  \label{fig:EdgeDiffConfs}
\end{figure*}

\begin{figure}[b!]
  \includegraphics[width=3.2in]{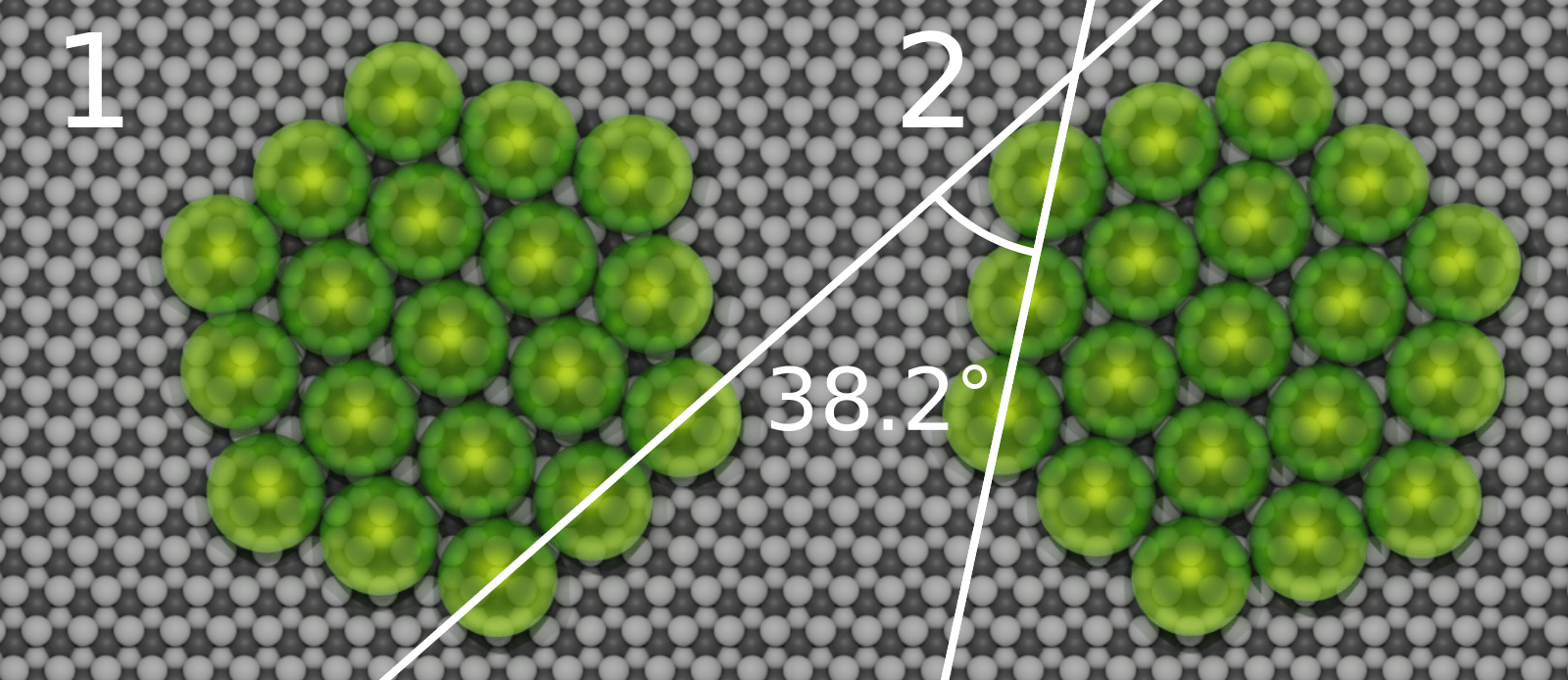}
  \caption{Two observed cluster orientations.}
  \label{fig:ClusterOrientations}
\end{figure}

After considering the dynamics of a single molecule on an otherwise empty substrate, we move on to transitions of a single molecule in the presence of other molecules. For the observation of edge diffusion, the simulation box size is increased to $\SI{80}{\angstrom} \times \SI{92}{\angstrom}$ to make space for small clusters consisting of a 19-membered hexagonal core cluster with additional adparticles placed at the clusters corners or edges.

To place the cluster in the correct initial orientation with respect to the substrate, we first run a few simulations where we initialize the cluster in an orientation parallel to the substrate and observe into which configurations it relaxes. As was also observed in experiments~\cite{LoskePaper}, the clusters can relax into two distinct orientations with an angle of $\SI{38.2}{\degree}$ between them (Fig.~\ref{fig:ClusterOrientations}). The adsorption sites are located on top of the third layer fluoride atoms, where the C$_{60}$ molecules are in close proximity to three first layer fluoride atoms. For the following simulations, we initialize all clusters with orientation 1 (Fig. \ref{fig:ClusterOrientations}).

\subsection{Modelling of the transition rates}

The goal of the following simulations is to determine a set of energy barriers and attempt rates for several types of transitions from observed transition rates. We estimate the transition rate for a given transition $i$ observed with interaction strength $\epsilon_\text{F}$, initial neighbours $n$ and at temperature $T$ using the unbiased estimator
\begin{equation}
  k_i(\epsilon_\text{F},n,T)=\frac{N_i(\epsilon_\text{F},n,T)-1}{t_\text{tot,i}(\epsilon_\text{F},n,T)},
\end{equation}
where $N_i(\epsilon_\text{F},n,T)$ is the number of observations of transition $i$ with $n$ initial neighbours at temperature $T$ and $t_\text{tot,i}(\epsilon_\text{F},n,T)$ is the total time we have observed an initial state that can go through the specified transition. As transition types $i$ we are considering edge diffusion transitions along A and B step edges ($i=$ ED-A/B), ascension to the second layer ($i=$ Asc) and dissociation from the cluster ($i=$ Diss). Assuming an Arrhenius type behaviour of the transition rates,
\begin{equation}
  k_i(\epsilon_\text{F},n,T)=\nu_{0,i}(\epsilon_\text{F},n)e^{-\Delta E_i(\epsilon_\text{F},n)/k_BT},
  \label{eq:Arrhenius}
\end{equation}
we can extract energy barriers $\Delta E_i(\epsilon_\text{F},n)$ and attempt rates $\nu_{0,i}(\epsilon_\text{F},n)$ from Arrhenius plots.

For the modelling of the energy barriers and attempt rates, we introduce a few constraints to ensure thermodynamic consistency and allow for interpolation between the values of $\epsilon_\text{F}$ that we have measured. As a general ansatz for the energy barrier of transition type $i$ with $n$ initial neighbours, we follow a bond counting approach and assume a linear dependence on the interaction parameter $\epsilon_\text{F}$,
\begin{equation}
  \Delta E_i(\epsilon_\text{F},n)=\Delta E_{i,0} + m_i\epsilon_\text{F}+(n-\hat{n})E_B,
  \label{eq:GenEnBarrs}
\end{equation}
with base energy $E_{i,0}$, slope $m_i$, and an effective binding energy that we set to $E_B=\SI{235}{\milli\electronvolt}$ (based on the results of Ref. \cite{Janke2020}). The parameter $\hat{n}$ in Eq.~\eqref{eq:GenEnBarrs} denotes the number of lateral bonds that can be sustained during the transition (in our case, $\hat{n}=0$ for dissociation, $\hat{n}=1$ for edge diffusion and $\hat{n}=2$ for ascension). For simplicity, we assume that the attempt rates $\nu_{0,i}(\epsilon_\text{F},n)$ for edge diffusion and dissociation are constants with respect to $\epsilon_\text{F}$ but can vary with transition type $i$ and number of initial neighbours $n$,
\begin{equation}
  \nu_{0,i}(\epsilon_\text{F},n)=\nu_{0,i}(n) \hspace{0.5cm} i\in\{\text{ED-A/B,Diss}\}.
  \label{eq:ConstNus}
\end{equation}
To ensure thermodynamic consistency of the model, we require the detailed balance condition
\begin{equation}
  \frac{k_\text{ED-A}(\epsilon_\text{F},2,T)}{k_\text{ED-A}(\epsilon_\text{F},1,T)}=\frac{k_\text{ED-B}(\epsilon_\text{F},2,T)}{k_\text{ED-B}(\epsilon_\text{F},1,T)}
  \label{eq:EdgeDiffRateAssumption}
\end{equation}
for the transition rates between A and B step edges. Inserting Eqs.~\eqref{eq:Arrhenius} and~\eqref{eq:GenEnBarrs}, we obtain the condition
\begin{equation}
\frac{\nu_{0,\text{ED-A}}(2)}{\nu_{0,\text{ED-A}}(1)}=\frac{\nu_{0,\text{ED-B}}(2)}{\nu_{0,\text{ED-B}}(1)}.
  \label{eq:EdgeDiffAttRateAssumption}
\end{equation}
for the attempt rates. Only in the case of the ascension transition are we using a linear fit function $\nu_{0,\text{Asc}}(\epsilon_\text{F},n)=y+m_\text{0,Asc}\cdot \epsilon_\text{F}$ to allow for the ascension barriers to be equal to separately measured potential energies $E_\text{Pot}$ (see Supplementary Information for justification~\cite{sm}).

\subsection{One initial neighbour}

\begin{figure}[t!]
  \includegraphics[width=3.2in]{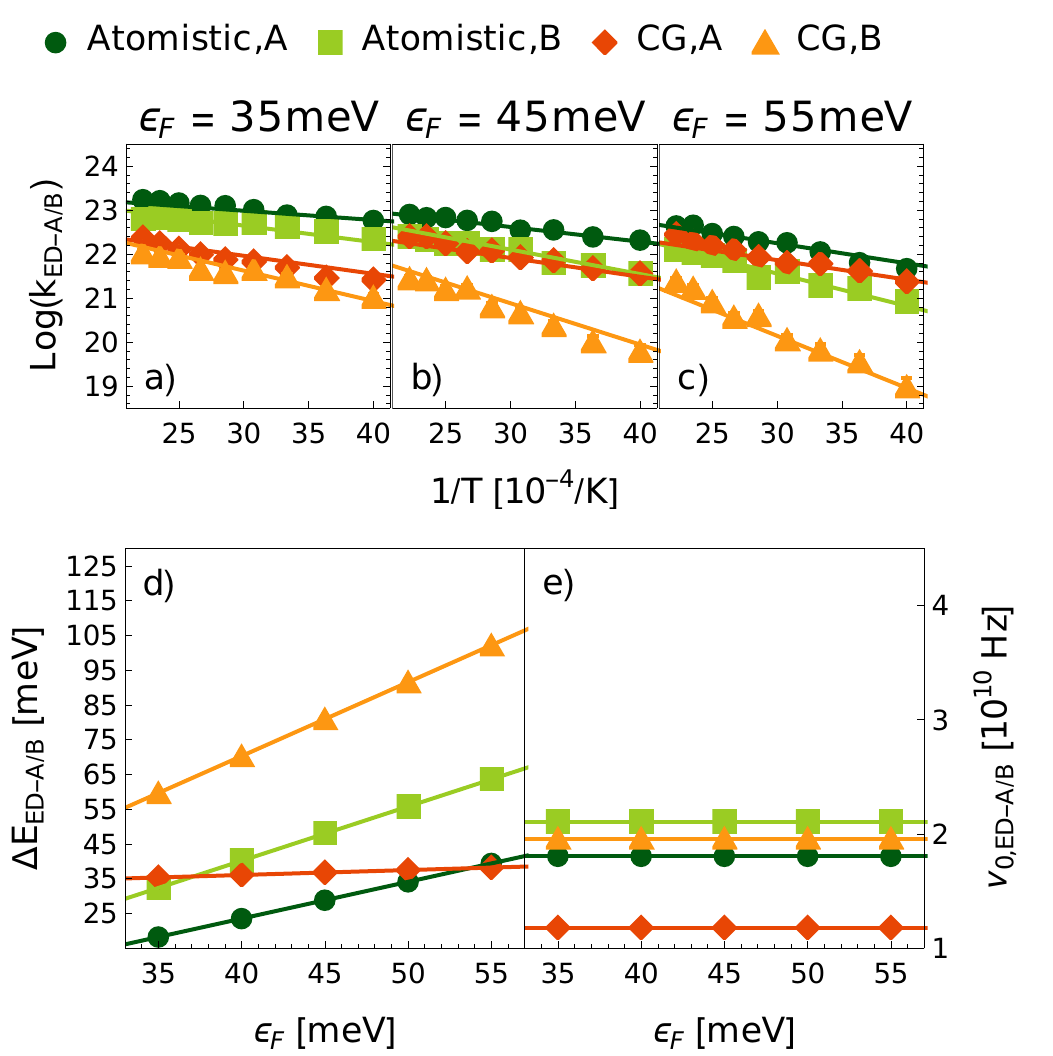}
  \caption{(a-c)~Arrhenius plots and extracted (d)~energy barriers and (e)~attempt rates for edge diffusion with one initial neighbour ($n=1$). Note that the functional dependence is constrained to linear through Eq.~\eqref{eq:GenEnBarrs}.}
  \label{fig:1NArrhenius}
\end{figure}

In the case of one initial neighbour, we put two additional C$_{60}$ molecules on opposing corners of the core cluster so that they are bound to a single core cluster molecule (Fig. \ref{fig:EdgeDiffConfs} configuration 1N). The simulations are initialized in a similar fashion as before, using a combination of Langevin thermostat and velocity rescaling to achieve random initial velocities at a given temperature $T\in [225,450]\SI{}{\kelvin}$. After the initialization, the simulation runs until one of the corner molecules leaves its initial position. The coordination number of the two corner molecules acts as a trigger to detect the transition since it changes from one to zero or two after a transition. After a transition has occurred, the simulation time is recorded and a few snapshots are taken in $\SI{1}{\pico\second}$ intervals to categorize the transition. To test if there is a preferred direction of edge diffusion (as is the case for C$_{60}$ on C$_{60}(111)$ diffusion), we categorize them into transitions towards A and B edges (Fig.~\ref{fig:EdgeDiffConfs}).

The resulting Arrhenius plots for the transition rates of A and B step edge diffusion with one initial neighbour are shown in Fig.~\ref{fig:1NArrhenius}(a-c). In these plots, we see that the atomistic model has an overall higher transition rate than the coarse-grained model. Both models show a preferred diffusion towards the A step edge that increases with $\epsilon_\text{F}$, with the difference between the A and B step transition becoming much larger in the coarse-grained model. The corresponding energy barriers and attempt rates are plotted in Fig.~\ref{fig:1NArrhenius}(d-e). Since edge diffusion with one and two initial neighbours is connected via Eqs.~\ref{eq:GenEnBarrs} and \ref{eq:EdgeDiffAttRateAssumption} in our modelling, this model also incorporates data from the following two neighbour simulations.

\subsection{Two initial neighbours}

\begin{figure}[b!]
  \includegraphics[width=3.2in]{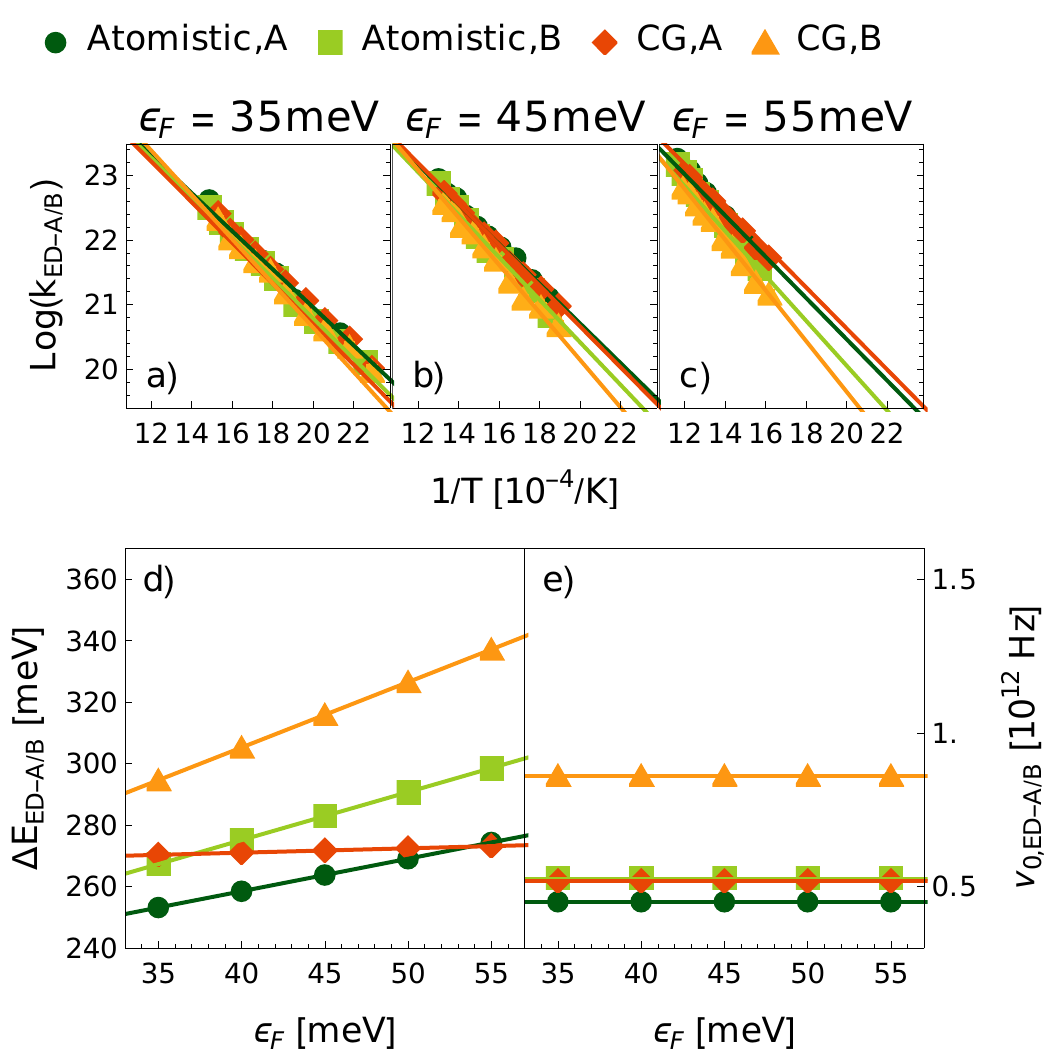}
  \caption{(a-c)~Arrhenius plots, (d)~extracted energy barriers, and (e)~attempt rates for edge diffusion with two initial neighbours ($n=2$). Note that the functional dependence is constrained to linear through Eq.~\eqref{eq:GenEnBarrs}.}
  \label{fig:2NArrheniusEdge}
\end{figure}

For the observation of elementary transitions with two initial neighbours, we have set up two distinct initial configurations. In configuration 2N1 (Fig.~\ref{fig:EdgeDiffConfs}), six C$_{60}$ molecules are positioned at the edges of the core cluster in a position where all the relevant transitions can be observed (namely ascension, dissociation, and edge diffusion with one or two final neighbours). Configuration 2N2 was set up to test for asymmetry with regards to the direction of the transition. However, in the end no asymmetry was observed (the transition rates for $A_1$ and $A_2$, as well as $B_1$ an $B_2$ in Fig. \ref{fig:EdgeDiffConfs} 2N2 were the same) and the results from configurations 1 and 2 were merged.

While the initialization process and the analysis remains unchanged, the detection mechanism of transition was adapted since the coordination number between the C$_{60}$ molecules does not necessarily change during a transition. Additional non-interacting ``dummy'' particles were placed at the initial locations of the adparticles and the coordination number between the adparticles and dummy particles was monitored. To be able to observe the rare ascension and dissociation transitions sufficiently often, we increase the temperature as high as possible without disintegrating the cluster. Because the stability of the cluster increases with $\epsilon_\text{F}$, we therefore vary the temperature range starting at $T\in [450,720]\SI{}{\kelvin}$ for $\epsilon_\text{F}=\SI{35}{\milli\electronvolt}$ and ramping up to $T\in [650,920]\SI{}{\kelvin}$ for $\epsilon_\text{F}=\SI{55}{\milli\electronvolt}$.

\begin{figure}[t!]
  \includegraphics[width=3.2in]{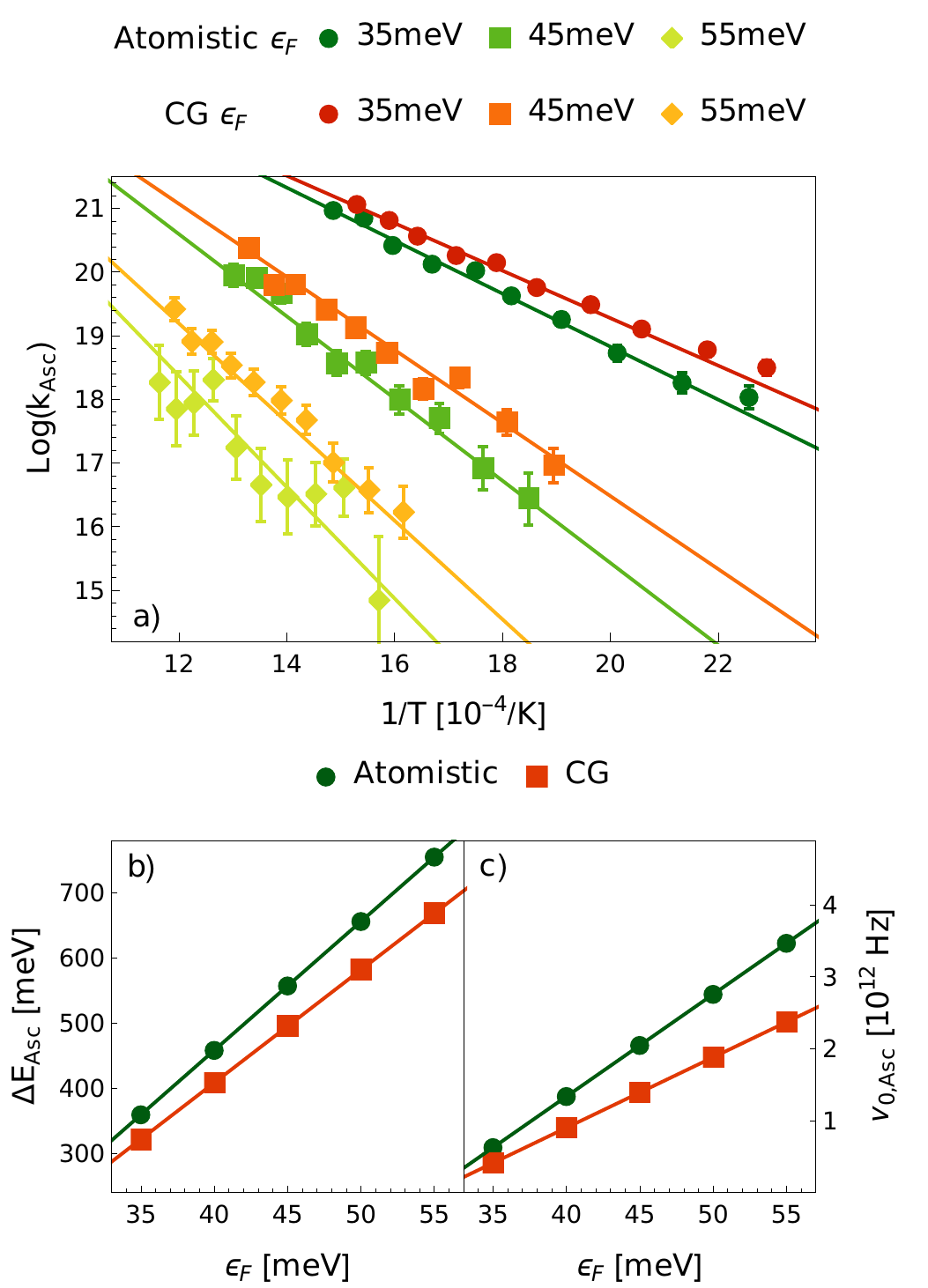}
  \caption{(a)~Arrhenius plots, (b)~extracted energy barriers, and (c)~attempt rates for ascension with two initial neighbours ($n=2$). Note that the functional dependence is constrained to linear through Eq.~\eqref{eq:GenEnBarrs}.}
  \label{fig:2NArrheniusAsc}
\end{figure}

For the case of edge diffusion, Arrhenius fits and the resulting parameter model are shown in Fig. \ref{fig:2NArrheniusEdge}. In contrast to the one-neighbour edge diffusion case, we now see only little difference between the atomistic and coarse-grained model. For low values of $\epsilon_\text{F}$, both models show no significant difference between the transition rates on A and B edges, while for $\epsilon_\text{F}\geq \SI{45}{\milli\electronvolt}$ a clear separation between the transition rates of the two edge types arises. The agreement of the two models in this case can be explained by two effects. Firstly, the ability of the rotational degrees of freedom to lower the effective energy barriers in the atomistic model may be diminished with higher coordination. Secondly, as we have also observed in Sec. \ref{sec:freediff}, the increased rotational diffusion in the higher temperature range may lead the atomistic model to more closely exhibit the rotationally averaged interaction of the coarse-grained model.

\begin{figure}[t!]
  \includegraphics[width=3.2in]{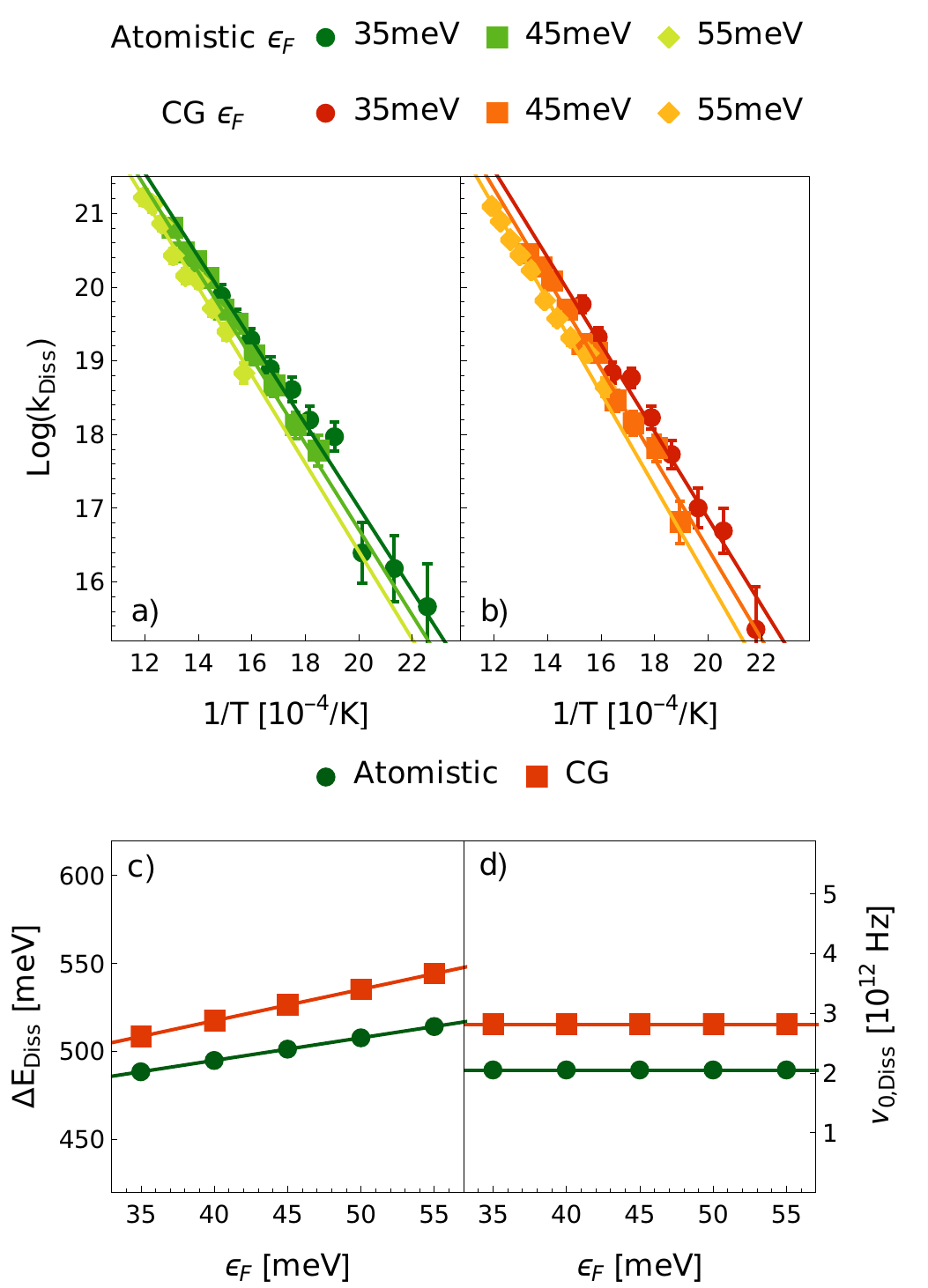}
  \caption{(a,b)~Arrhenius plots, (c)~extracted energy barriers, and (d)~attempt rates for dissociation with two initial neighbours ($n=2$). Note that the functional dependence is constrained to linear through Eq.~\eqref{eq:GenEnBarrs}.}
  \label{fig:2NArrheniusDiss}
\end{figure}

The Arrhenius plots and parameter model for the ascension transition are shown in Fig. \ref{fig:2NArrheniusAsc}. In the Arrhenius plots we can see that for the same value of $\epsilon_\text{F}$, the coarse-grained model shows a higher ascension rate, which is in accordance with the fact that the adparticle-substrate binding energy is weaker in this model (Fig.~\ref{fig:MEPvsPPD}). The energy barriers of the model [Fig.~\ref{fig:2NArrheniusAsc}(b)] are set to match the average potential energy of a single C$_{60}$ adparticle on the substrate $E_\text{pot}$, which were separately measured (see Supplementary Information~\cite{sm}). Finally, the transition rates for dissociation turn out to only weakly depend on $\epsilon_\text{F}$ and are very similar for both models [Fig.~\ref{fig:2NArrheniusDiss}(a,b)]. The derived energy barriers and attempt rates are shown in Fig.~\ref{fig:2NArrheniusDiss}(c,d).

\section{Conclusions}

In this paper, we have performed and analysed molecular dynamics simulations of free and edge diffusion processes of C$_{60}$ molecules on CaF$_2(111)$ using two distinct descriptions for the adparticle-substrate interaction. We compare an atomistic rigid body model with Lennard-Jones C-F and Buckingham type C-Ca interactions to a coarse-grained model with central-body C$_{60}$-F and C$_{60}$-Ca interactions, which we have derived in analogy to the well-known Girifalco potential~\cite{GirifalcoPot,GirifalcoPot92}. Force-field parameters are taken from the literature except for the fluoride-carbon energy parameter $\epsilon_\text{F}$, which we leave as a single free parameter to tune the substrate binding energy.

We find that the two models can produce significantly different results, which are especially apparent in the minimum energy paths (Sec.~\ref{sec:MEPs}) and for the free diffusion of a single molecules at low temperature. These differences can be attributed to a change in the rotational diffusion of the atomistic model at around $\SI{163(1)}{\kelvin}$, which turns out to be independent of $\epsilon_\text{F}$. It is accompanied by a crossover of the translational diffusion in a temperatures range $100-\SI{200}{\kelvin}$ now depending on the value of $\epsilon_\text{F}$. Since the rotational degrees of freedom have been integrated out for the coarse-grained molecular model, it is lacking these crossovers but agrees quantitatively with the high-temperature behavior of the atomistic model [Fig.~\ref{fig:enBarrSummary}]. Our results agree with the idea of a ``lock-in'' of the molecular orientation that is overcome at high temperatures, where the coarse-grained model becomes applicable. The free diffusion barriers extracted from the minimum energy paths and molecular dynamics calculations of both models ($15-\SI{90}{\milli\electronvolt}$) remain at variance with the experimental value of $\SI{214(16)}{\milli\electronvolt}$ that was calculated indirectly from cluster densities of deposition experiments~\cite{LoskePaper}.

\begin{table}[t!]
\caption{Final model parameters derived from our MD simulations. Listed are the parameters of Eq.~\eqref{eq:GenEnBarrs} together with the attempt frequencies. The values for free diffusion (FD) are derived from the MSD calculations of Sec.~\ref{sec:freediff} (for more details, see Supplementary Information~\cite{sm}).}
\label{tab:modelParams}
\centering
\begin{tabular}{ccc|c|c|c}
  \hline\hline
  Model & $i$ & $n$ & $\Delta E_{i,0}[\SI{}{\milli\electronvolt}]$ & $m_i$ & $\nu_{0,i}(n)[\SI{}{\giga\hertz}]$\\
  \hline
  \multirow{6}{*}{Atom.}& FD-LT & 0 & $-8.07$ & $0.6546$ & $153$ \\
  & FD-HT & 0 & $-45.33$ & $2.664$ & $9025$ \\
  & ED-A & 1 & \multirow{2}{*}{$-18.9$} & \multirow{2}{*}{$1.06$} & $18.0$ \\  
  & ED-A & 2 &  &  & $450$ \\
  & ED-B & 1 & \multirow{2}{*}{$-22.4$} & \multirow{2}{*}{$1.56$} & $21.0$ \\
  & ED-B & 2 &  &  & $525$ \\
  & Diss & 2 & $-26.5$ & $1.29$ & $2050$ \\
  & Asc & 2 & $-333$ & $19.8$ & $-4355+142\frac{\SI{}{\giga\hertz}}{\SI{}{\milli\electronvolt}}\epsilon_\text{F}$ \\
  \hline
  \multirow{6}{*}{CG}& FD & 0 & $-68.23$ & $2.896$ & $27301$ \\
  & ED-A & 1 & \multirow{2}{*}{$30.3$} & \multirow{2}{*}{$0.142$} & 11.8 \\  
  & ED-A & 2 &  &  & 518 \\
  & ED-B & 1 & \multirow{2}{*}{$-14.8$} & \multirow{2}{*}{$2.13$} & 19.6 \\
  & ED-B & 2 &  &  & 861 \\
  & Diss & 2 & $-23.8$ & $1.78$ & $2816$ \\
  & Asc & 2 & $-285$ & $17.3$ & $-3025+98\frac{\SI{}{\giga\hertz}}{\SI{}{\milli\electronvolt}}\epsilon_\text{F}$ \\
  \hline\hline
\end{tabular}
\end{table}

The second central result of our work is a comprehensive set of transition rates to be used in kinetic Monte Carlo simulations. To this end we have systematically probed edge diffusion transitions with one and two initial neighbours. While the two models still show significantly different behaviour for edge diffusion with one initial neighbour, the transitions with two initial neighbours start to look more similar, presumably because of the higher coordination and higher temperature at which the simulations were performed. The parameters needed to obtain the energy barriers in Eq.~\eqref{eq:GenEnBarrs} are listed in Tab.~\ref{tab:modelParams} together with the attempt frequencies. Also included in this table are model parameters for the free diffusion transition (FD), the modelling plots for which can be found in the Supplementary Information~\cite{sm}. In the atomistic case, the free diffusion parameters are separated into low temperature (FD-LT) and high temperature (FD-HT) regime. A comprehensive analysis of the resulting morphologies will be presented in a forthcoming publication. Our approach demonstrates how the problem of building a reasonable KMC model with a low number of free parameters can be tackled and solved in an efficient and systematic way with the use of MD simulations of individual elementary transitions. In the future second part we are going to use these KMC models in extensive KMC simulation to reproduce experimentally observed cluster morphologies.

\section*{Acknowledgements}

We thank A. K\"uhnle and her group members for stimulating discussions. We acknowledge funding from the Deutsche Forschungsgemeinschaft (Grant No. 319880407). All numerical computations were performed on the MOGON II Cluster at ZDV Mainz.

\section*{Data availability}

The data that support the findings of this study are available from the corresponding author upon reasonable request.

\section{Appendix}

\subsection{Rotational Diffusion Coefficient}
\label{app:rotDiff}

To determine the rotational diffusion coefficient $D_r$, we use a definition of the mean squared angular displacement (MSAD) analogous to the translational case,
\begin{align}
\text{MSAD}(t)=\mean{(\theta (0)-\theta (t))^2}=2f_rD_rt,
\label{eq:MSAD}
\end{align}
with $f_r=3$ the number of rotational degrees of freedom. To determine the MSADs, we measure the angular displacements $(\theta (0)-\theta (t))$ as the angle of the optimal rotation between the two conformations of the C$_{60}$ molecule at times $0$ and $t$. The optimal rotation is determined with a quaternion approach as described in \cite{Karney2007,Liu2009}. For sufficiently small times $t$ we find that the squared angular displacements (SADs) follow an exponential distribution,
\begin{align*}
p_t^\star(\text{SAD})=\frac{1}{\text{MSAD}(t)}\cdot e^{-\text{SAD}/\text{MSAD}(t)}.
\end{align*}
For larger values of $t$ the distribution changes because the SAD of an optimal rotation can only take values between $\text{SAD}=0$ and $\text{SAD}=\pi^2$. Effectively, the exponential distribution gets reflected back and forth between those two values and the resulting distribution can be derived to be
\begin{align}
p_t(x)=\frac{2\lambda \cosh{(\lambda x)}}{1-e^{-2\lambda \pi^2}}-\lambda e^{\lambda x},
\label{eq:SADDistrib}
\end{align}
where we have substituted $\lambda=1/\text{MSAD}(t)$ and $x=\text{SAD}$ for better readability. Using Eq.~\eqref{eq:SADDistrib} as a fit function to the distribution of SADs, we obtain the values of MSAD(t). From the MSADs we then determine the rotational diffusion coefficients $D_r$ via Eq.~\eqref{eq:MSAD}.

%

\end{document}